\input harvmac

\Title {UCD-PHY-95-39}
{{\vbox {\centerline{Time-Symmetric Initial Data  }\break
\centerline{
 for Multi-Body Solutions in 
Three Dimensions}}}}
  \bigskip 
\centerline{Alan R. Steif}
 \bigskip\centerline{\it Department of
Physics}
\centerline {\it   University of California }
  \centerline{\it Davis, CA 95616 }
\centerline{\it steif@dirac.ucdavis.edu}
  \vskip .2in

\noindent
ABSTRACT: Time-symmetric initial data  for two-body  solutions in 
three dimensional anti-deSitter gravity are found.  The spatial geometry has constant negative curvature and   is constructed
as a quotient of two-dimensional  hyperbolic space.  Apparent horizons correspond to closed
 geodesics.  In an open universe, it is shown that
two black holes cannot
exist separately, but 
are necessarily enclosed by a third horizon.   
In a closed universe, two separate black holes can exist
provided there is an additional image mass. 

\Date {}
\def\S{\Sigma}
\def\Ads3{three dimensional anti-deSitter space}

\def\G{\Gamma}
\def\ra{\rightarrow}

\def\su{$SU(1,1)$}

\def\a{\alpha}
\def\b{\beta}

\def\ra{\rightarrow}

\def\m{\mu}

\def\n{\nu}

\def\D{\bf d}

\def\D3{\hbox{$D_3\kern -10pt / \kern 10pt$}}
\def\Aslash{\hbox{$ A^a\kern -10pt / \kern +10pt$}}
\def\paslash{\hbox{$ \partial \kern -6pt / \kern +6pt$}}

\def\r{\rho}

\def\tphi{\tilde\phi}

\def\tr{\tilde r}

\newsec{Introduction}

 Black holes in three spacetime dimensions  {\ref\btz{
M. Banados, C. Teitelboim, and J. Zanelli, {\it Phys. Rev. Lett.} {\bf 69}
(1992) 1849;  M. Banados, M. Henneaux,  C. Teitelboim, and J. Zanelli,
 {\it Phys. Rev. D} {\bf 48} (1993) 1506.}} {\ref\sc{For a review, see
R. Mann, ``Lower Dimensional Black Holes: Inside and Out'',
Proceedings  of the Winnipeg Conference on Heat Kernels
and Quantum Gravity; 
S. Carlip, ``The $2+1$ Dimensional Black Hole'', to appear in {\it Class. Quant. Grav.}}}   share many of   the features of four dimensional black holes. 
In this paper, the issue of constructing multi-body and in particular multi-black hole solutions is considered.
 In {\ref\oh{O. Coussaert and M. Henneaux, in  {\it {  Geometry of Constrained Dynamical Systems}}, ed. John M. Charap, Cambridge University Press, 1995.},}
 it was shown that  there are no static multi-black hole solutions in
 $2+1$ dimensions. Indeed, since there is a negative cosmological constant present, one would expect the black holes to attract and only time-dependent  solutions to exist. In {\ref\clement{ G. Clement, {\it Phys. Rev. D} 
{\bf D50} (1994) R7119.}}, it was  shown that additional conical singularities will appear in the time-dependent solutions. To better understand these multi-body solutions, we focus on 
the problem of constructing 
 initial data for two bodies initially at rest. We take advantage of
 the fact that the space exterior to the sources has constant negative 
curvature and therefore can be constructed as a quotient of hyperbolic space.

\newsec{Time-Symmetric Initial Data in Three Dimensions}

 Black holes  in $2+1$ dimensions   are   solutions to Einstein's equations
with a negative cosmological constant, $\Lambda$,
\eqn\einsteineqn{
G_{\m\n} +\Lambda g_{\m\n} = 8\pi G T_{\mu\nu} ,\quad \Lambda <0
.}
The initial data constraints for {\einsteineqn} 
on an initial spacelike slice $\Sigma$
with spatial metric $h_{ij}$ and      extrinsic curvature $K_{ij}$ are given
by 
\eqn\constraints{\eqalign{
{R\over 2} + K_{ij} K^{ij} - K^2 + &{1\over l^2} = 8\pi G T_{\mu\nu} n^{\mu} n^{\nu}, \quad l^2 = -\Lambda^{-1}, \quad K\equiv K^i_{\; i} \cr
\nabla _j K^j_{\,i} & - \nabla_i K = 8\pi G T_{i \mu} n^{\mu} \cr
}}
where $n^{\mu}$ is the normal to $\S$ and  $R$ is 
the scalar curvature of $h_{ij}$.      $i,j, \cdots$ refer 
to indices tangent to the
spatial slice.
In this paper, we are concerned with the case $ K_{ij} =0$ corresponding
to time-symmetric or  momentarily static 
   initial data.  
The momentum constraint is satisfied if $T_{i \mu} n^{\mu} =0 $ while  the Hamiltonian   constraint becomes
 \eqn\Ham{ {R\over 2} + {1\over l^2} = 8\pi G \rho ,\quad 
\quad \rho = T_{ \mu\nu} n^{\mu} n^{\nu} 
.}

 On an apparent horizon, $S$, the convergence (or expansion) of outgoing null geodesics vanishes. 
In terms of $h_{ij}$ and $K_{ij}$, this corresponds to the condition
\eqn\apparenthorizon{
 H = (h^{ij} - {\tilde n}^i {\tilde n}^j ) K_{ij}
} where $H$ is the mean spatial curvature of $S$ viewed as a surface 
imbedded
in the $D-1$ dimensional space $\Sigma$ with metric $h_{ij}$
and where ${\tilde n}^i$ is the normal to $S$ in $\Sigma$.
 For time-symmetric initial data, $S$ is an apparent horizon if $H=0$, 
{\it i.e.} if $S$ is a minimal surface. 
A curve which is minimal is a geodesic.  Hence, apparent horizons for time-symmetric
initial data in $2+1$ dimensions are closed geodesics.

 \newsec{Initial Data for   
Static Circularly Symmetric One-Body Solutions  }

  We first consider initial data for the static one-body solutions.  
  The  static  circularly symmetric solutions to {\einsteineqn} 
     are  given by {\btz}
\eqn\bhmetric{
dS^2  = - ({  r^2\over l^2} -8GM) d t^2 + ({  r^2\over l^2} -8GM)^{-1} d  r^2 +
 r^2 d  \phi^2
, \quad   0 <  \phi < 2\pi
}
where   $M$ is the total mass.
For various ranges of $M$, {\bhmetric} describes the following solutions:

(1) $M>0$:  black hole  with   event horizon  located at $  r_H =
 ({8GM})^{1/2}l$ and singularity at $r=0$

(2) $M=0$: black hole vacuum

(3) $-{1\over 8G} < M < 0:$  one-particle solutions with a naked conical singularity at
$r=0  $ and no event horizon {  {\ref\djt{
S. Deser, R. Jackiw, and G. 't Hooft, {\it Ann. Phys.} (NY) {\bf 152} (1984) 220.}} \ref\dj{S. Deser and R.
Jackiw, {\it Ann. Phys.} {\bf 153} (1984) 405.}}

(4) $M= -{1\over 8G}:$ three dimensional  anti-deSitter space.

 The $t=0$ initial spacelike slices of {\bhmetric} are time-symmetric 
and hence,  from {\Ham},  have constant negative curvature. This  implies that
they   can be obtained as   quotients of  two-dimensional hyperbolic space.
 We now review this construction.

\subsec{ Two-Dimensional Hyperbolic Space }
  
  Two-dimensional  hyperbolic space, $H^2$, can be described as  the two-dimensional hypersurface
\eqn\hypersurface{
T^2 -X^2 -Y^2 = l^2 
}
in the flat three-dimensional space with metric of signature $(-++):$
\eqn\flat{
dS^2 = -dT^2 +dX^2 + dY^2
.}
 The Poincare disk model  for hyperbolic space is 
\eqn\disk{
ds^2  = {4\over( 1 - z\bar z/l^2)^2 }dz d\bar z,\quad    \quad  0\leq |z|< l,    \quad z= \r e^{i\phi} ,
}
 and can be obtained by the  stereographic projection of    the hypersurface {\hypersurface} 
   through the point $(-1,0,0)$ in
the $(T, X, Y)$ space  onto the disk of radius
$l$ in the $X-Y$ plane.  
The boundary $|z|=l$ in {\disk} is  spatial infinity.
 Geodesics on the Poincare  disk are segments of circles or lines which
intersect the boundary of the disk orthogonally.
The isometry group of {\disk} is $SU(1,1)$ with action
\eqn\suaction{ 
z/l \ra  { \a (z /l)+ \b  \over  \bar\b (z /l)+\bar\a },\quad |\a|^2 -|\b |^2 =1
.}

  Another representation of $H^2$ which will be useful is  the Poincare metric on the upper half $xy$-plane: 
  \eqn\upperhalfmetric{
ds^2 = l^2({dx^2 +dy^2\over y^2}),\quad y>0  
.}
This  can be obtained from  {\flat}  by the imbedding
\eqn\imbedxy{
T+Y = l^2/y, \quad T-Y = {x^2+y^2\over y},\quad X =  {x\over y} l .
}
It  can   be obtained from {\disk} by applying an inversion in a circle in the $z$ plane given by 
$z/l \ra (z/l -i)(-iz/l+1)^{-1}$ where $z= x+iy$.  Geodesics on the  upper half plane are vertical lines or semi-circles which intersect the real axis
orthogonally. 
The isometry group of   {\upperhalfmetric} is $SL(2,R)$ with action
\eqn\slaction{ 
z/l \ra  {  a (z/l ) +b  \over  c (z/l)  +d  },\quad 
{\pmatrix{a&b\cr c&d\cr}} 
\in  SL(2,R)    .}
  $ SL(2,R)  $   and    $ SU(1,1)$ isometries are related by conjugation
 \eqn\conjugation{
\tilde S = N S N^{-1},\;\;  N= {1\over {\sqrt{2}}}{\pmatrix{1&-i\cr -i& 1\cr}}
,  \quad S\in SL(2,R), \quad \tilde S \in SU(1,1)
.
}
  Finally, in terms of polar coordinates   $r= 2{\r\over 1- \r^2 /l^2}$,  {\disk}   takes the form 
\eqn\adsid{
ds^2 = {dr^2\over r^2/l^2 +1} + r^2 d\phi^2,\quad r\geq 0 ,\quad 0\leq \phi < 2\pi
.}
This is the $t=0$ spatial geometry of  three dimensional anti-deSitter space 
  {\bhmetric}  with $M= -1/(8G)$ .
  
   Let us briefly discuss the conjugacy classes of isometries of $H^2$.
   Two isometries are conjugate if and
only if the traces of the corresponding  $SL(2,R)$ or $SU(1,1)$ matrices, $\Gamma$, are equal. There are three types of conjugacy 
classes    :
{\it elliptic}  $((Tr\G )^2  < 4)$, {\it hyperbolic} $((Tr\G )^2    >  4)$ or {\it parabolic} $((Tr\G )^2    = 4)$. Elliptic isometries
are conjugate to rotations about the origin in the Poincare disk. Hyperbolic
  isometries are conjugate to
scalings  in the upper half plane.  There is only one parabolic conjugacy class 
corresponding to  isometries which are conjugate to  a   translation in the
 $x$-direction in the upper-half plane.  Elliptic isometries  have one fixed  point,   hyperbolic isometries have  two fixed points 
both at infinity, and parabolic  
isometries have  one  fixed  point at infinity. The $t=0$ initial slices for the
 one-body solutions {\bhmetric}
    can be obtained
essentially by identifying $H^2$ periodically in an isometry generator.  
 Elements conjugate to one another generate
isometric quotients. 
As we   shall see below, elliptic isometries generate the $M<0$ one-particle
solutions, hyperbolic isometries $M>0$ black holes, and the parabolic
conjugacy class generates the $M=0$ black hole vacuum.

  \subsec{$M>0$ Black Hole   Solutions   }
  
The spatial geometry of the black hole solution can be obtained
by identifying $H^2$ periodically in  a hyperbolic generator  
{\btz}.   The  $t=0$ spatial slice of the   black hole spacetime {\bhmetric}   is given by 
\eqn\spatialbh {
ds^2  =  ({  r^2\over l^2} - 8GM)^{-1} d  r^2 +
 r^2 d  \phi^2
,\quad       0 <  \phi < 2\pi .
}
Defining the radial coordinate $x = (r^2 - 8GMl^2)^{1/2}$,
one obtains 
\eqn\wormhole  {
ds^2  =  ({  x^{ 2}\over l^2} + 8GM)^{-1} d  x^{ 2} +
 ( x^{ 2}  + 8GMl^2) d  \phi^2
,\quad    -\infty < x < \infty , \quad    0 <  \phi < 2\pi .
}
The complete spatial geometry   has a wormhole structure with cylindrical topology,   $S^1\times R$. It is    similar to the Einstein-Rosen bridge in the  Schwarzschild solution except it
is not asymptotically flat. 
 The wormhole mouth is at the horizon $r_H = (8GM)^{1/2} l$.  

The simplest way to see how {\spatialbh} can be obtained from a quotient of 
$H^2$ is using    the
coordinates $( \tr, \tphi )$ 
defined by the imbedding {\hypersurface}
\eqn\trphicoor{ 
T = \tr \cosh \, \tphi,  \quad    \;\quad X   =
  \sqrt{\tr^2 - l^2 }   ,      
      \; \quad Y  = \tr \sinh \, \tphi ,   \quad\quad  \tphi \in ( - \infty, \infty ) ,   \quad\quad    l < \tilde r < \infty  
 . }
Note that $\tphi $  has infinite range because it 
 parameterizes {\it boosts} in the $T - Y$ plane.
 Inserting   into {\flat},  the metric becomes
\eqn\trpmetric{
ds^2  =  ({\tr^2\over l^2} -1)^{-1} d\tr^2 +
\tr^2 d \tphi^2
, \quad   \tphi \in ( - \infty, \infty ) .
}
  If one identifies
$\tphi$  with period $2 ({8GM})^{1/2} \pi$ where $M$ is
 the mass of
the black hole, and rescales the coordinates
\eqn\rescale{
\tr = r/ ({8GM})^{1/2}  \quad   \tphi =  ({8GM})^{1/2}\,\phi
,
}
one obtains the black hole spatial geometry {\spatialbh}.

 In terms of the coordinates {\imbedxy}  $(x,y)$
on the upper half-plane,
 the identification $\tphi \sim \tphi + 2 ({8GM})^{1/2} \pi$   corresponds
to the identification by the scaling  $(x,y) \sim  e^{-2({8GM})^{1/2}\pi} ( x,   y)$.
  The black hole spatial geometry   is then  the semi-annulus in the upper half-plane
  with outer unit radius   and inner  radius $e^{-2 ({8GM})^{1/2}\pi} $ identified 
{{\ref\scteit{
S. Carlip and C. Teitelboim, {\it Phys. Rev. D } {\bf 51} (1995) 622.}}}. 
 From {\conjugation}, the corresponding $SU(1,1)$ matrix 
generating the identifications is given by 
\eqn\blackholemat{
 \Gamma_M = 
\pmatrix{\cosh ({\pi (8GM)^{1/2}}) & i\sinh ({\pi (8GM)^{1/2}})   \cr
 -i\sinh ({\pi (8GM)^{1/2}})  &  \cosh ({\pi (8GM)^{1/2}})\cr}
 .}
Hence, the $M>0$ black hole is obtained by identifying two geodesics which do not intersect and are not tangential to one another at infinity.
 
 \subsec{$ -1/(8G) < M <0 $ Solutions  as    Quotients  of 
Hyperbolic Space}

The $ -1/(8G) < M <0 $  solutions \eqn\particlemetric{
ds^2  =    ({  r^2\over l^2}+ (1-4Gm)^2)^{-1} d  r^2    +
 r^2 d  \phi^2, \quad m = {1\over 4G} - ({-M\over 2G})^{1/2} 
}
 are  the familiar anti-deSitter
conical spaces with a particle source {\djt} 
obtained by excising a wedge
of deficit angle $8\pi Gm$ from the Poincare disk  with  vertex located  at the  source.   The edges of the wedge are   then identified by  a rotation     $z \ra e^{i 8\pi Gm} z$ which  as an element of {\su} 
is given by
\eqn\particlemat{
\Gamma_m  = \pmatrix{ e^{i 4\pi Gm}& 0 \cr 
0  & e^{-i 4\pi Gm}\cr }
.}
$m$ is the proper mass defined
by $m = \int \rho dA$.       The maximum allowable deficit angle of $2\pi$  corresponds to $m_c \equiv {1\over 4G}$. For a particle located at a different point,  the wedge excised   consists of two geodesic edges with  the same deficit angle 
  since the model is conformal.

  \subsec{$M=0$ Black Hole  Vacuum}
   The  $t=0$ spatial geometry of the $M=0$ 
    black hole vacuum  is given by 
\eqn\mzero{
ds^2  =   l^2{d  r^2\over r^2} +
 r^2 d  \phi^2
.}
This is  a   surface of revolution
of constant negative curvature 
known as a {\it tractroid}.
 Transforming to the new coordinates $y = l /r,\;\; x =\phi $,
{\mzero} becomes
\eqn\mzerob{
ds^2 = l^2 ({dx^2 +   dy^2\over y^2})
, \quad y >0,   \quad 0\leq x < 
 2\pi 
}
Comparing with {\upperhalfmetric}, the $M=0$ spatial geometry is the upper half plane with $x$ identified
 periodically in 
$2\pi .$
 Alternatively, using  {\conjugation} it can be obtained as a quotient of the Poincare disk by the 
$SU(1,1)$  parabolic transformation
\eqn\vacuummat{ \Gamma_{M=0} = 
\pmatrix{ 1 +{\pi}i  &  {\pi} \cr {\pi}  & 1-  {\pi}i\cr}
\in SU(1,1)  
}
     Identifying in $x$ with   different periods are conjugate
transformations. Hence, the resulting spaces generated by   identifying in
$x$ with different periods are  
all  isometric to the $M=0$ solution.
The $M=0$ solution is obtained by identifying two geodesics which are
 tangential to
one another at ${\infty}$.

The generator for the $M=0$ solution {\vacuummat} can be obtained  by   conjugating the generator for   the $M<0$ conical  solution {\particlemat}
  by a  
translation $T_d$ by a distance $d$ in the $x-$direction in the Poincare disk
\eqn\translation{
T_d = \pmatrix{ \cosh (d/2l)    & \sinh  (d/2l) \cr 
\sinh  (d/2l)    & \cosh  (d/2l)\cr}
} 
and taking the  simultaneous limit $m\ra 0,\;\; d\ra \infty$ 
\eqn\contraction{
\Gamma_{M=0} = \lim_{m\ra 0, \; d\ra \infty}
{ T_d \Gamma_m T^{-1}_d ,\quad  
}}
with $m\cosh d/l = m_c$ fixed. This   limit is analogous to a contraction where a lightlike solution is obtained by taking a simultaneous
$m\ra 0,\; v\ra c $ limit with the energy held fixed.
The $M=0$ generator can   be obtained in the same way
from   the $M>0$ black hole generator {\blackholemat}.

\newsec{Two-Body Solutions}

In this section, we construct initial data for  two-body solutions. As 
in the one-body case, we obtain these solutions  
  by taking quotients
of $H^2$. 
While the quotient group for the one-body solutions is generated by one-element,
 for the multi-body solutions more generators are required. The total 
mass of the system, $M$, can be obtained from
the spatial metric at large distances. However, a simpler way is to
 obtain it from the generator
  for the system.  The  generator $\Gamma$ for the total system is the composition of 
the generators for the individual bodies.
Express $\Gamma$ as an effective one-body generator 
up to a conjugation corresponding to an overall isometry 
\eqn\effective{
\G  =    T \G_{M} T^{-1}
}
where $\G_M$ is given in {\blackholemat}.
The total mass can then be obtained from the trace:
\eqn\totalmass{
\cosh ( \pi (8GM)^{1/2})  = {1\over 2} Tr\, \Gamma .
}
  We now consider three two-body systems:
 two particles,  a particle and black hole,   and two black holes.
Depending on the masses and locations of the bodies, there are three qualitatively different
kinds of solutions: the space exterior to the two bodies is
open without horizons; the space exterior to the two bodies is
open with the two bodies enclosed by
an apparent horizon; or, the space exterior to the two bodies is  closed with additional image masses.
For some values of these parameters, there may be no solution (at least with
positive energy matter). The same kinds of solutions were found
in the case of a ring of pressure-free  dust 
{\ref\ps{Y. Peleg and A. Steif, {\it Phys. Rev. D} 
{\bf 51} (1995)  R3992.}}.

\subsec{Two-Particle Solutions}

Consider   two particles  initially at rest with   masses $m  $, $\tilde m$ 
 separated
by a geodesic distance $d$.  To construct the spatial geometry, 
it is convenient to use the Poincare disk representation.
   As in the flat case {\djt},
 we excise a wedge  for each of the particles with deficit angles $8\pi G m$ and 
$8\pi G \tilde m$. 
  For a particle of mass $m$
translated by a distance $d$, the generator is
\eqn\translatedparticle{
\G = T_d \G_m T_d^{-1} }
where $\G_m$ is the one-particle generator  {\particlemat} and 
$T_d$ is the translation  {\translation}.  The effective one-body
generator for  the whole system is then the product 
\eqn\twoparticlemat{
\G  =   \G_{m} T_d \G_{\tilde m} T_d^{-1}.
}
From {\totalmass}, the total mass 
of the system is given by
\eqn\twoparticlemass{
\cosh ( {\pi (8GM)^{1/2}}) =   
       \cosh (d/l) \sin  (4\pi G m)
  \sin  (4\pi G \tilde m) -\cos   (4\pi G m )    \cos   (4\pi G \tilde m)    .
 }

Depending on the values of the parameters $(m, \tilde m,d)$,  the space can be
 open
without a horizon, open with the two particles enclosed by a horizon, or closed 
with an additional  image mass.
If $m+ \tilde m < m_c$, the space is open with total mass given by
{\twoparticlemass}.  If $\tilde m =0$, we recover the one-particle mass $M = - (1-4Gm)^2/ (8G)$ from {\particlemetric}.
When $d=0$, we recover the flat space formula
$m_{total} =  m+ \tilde m$ where $M = - (1-4Gm_{total})^2/ (8G)$.
$M$ increases with $d$ due to the attractive anti-deSitter force, and increases with
the masses $m, \tilde m$ due to the rest mass contribution to the total energy.

Now consider the case $m+ \tilde m \geq m_c$. Recall that this condition
 means that the sum of the  deficit angles of the particles exceeds $2\pi$.
 In flat space, this would imply that
the universe closes with an additional image mass appearing. In anti-deSitter
 space,
 this is not necessarily the case. Consider the two subcases     $m  \leq
 m_c/2, \; 
\tilde m > m_c/2$ (or equivalently, $m  > m_c/2$ and $\tilde m \leq  m_c/2$) and 
 $m, \tilde m > m_c/2$ separately.

For  $m  \leq m_c/2$ and $\tilde m > m_c/2$, the different kinds of solutions
that can occur as $d$  decreases are as
follows. For 
$d$ in the range  
\eqn\twopartopen{
\cosh (d/l) >   {  \tan (4\pi G m)\over      - \tan (4\pi G \tilde m )}
 ,{\rm Open, \; No \;Horizon  }}
the space is open with total mass {\twoparticlemass} and  without any horizon. For $d$ in the range 
\eqn\twopartmerged{
f_c (m, \tilde m)  < \cosh (d/l)  \leq  
 {  \tan (4\pi G m)\over      - \tan (4\pi G \tilde m )} \quad 
 {\rm Black\; Hole }
}
with 
\eqn\fdef{
f_c (m, \tilde m) \equiv 
{1 +  \cos  (4\pi G m)  \cos  (4\pi G \tilde m) \over   
   \sin  (4\pi G m) \sin  (4\pi G \tilde m) },
}
the two particles are surrounded by a horizon. The size of the horizon is that of a black hole of mass $M$   
{\twoparticlemass} as can be verified  directly  from the geometry.
The space outside the horizon is of course  the black hole spatial geometry.
When $  \cosh (d/l) =   f_c (m, \tilde m) $, {\twoparticlemass} yields
$M=0$, and the space outside the particles forms
an  infinite thin throat at  infinity which is identical to the $M=0$ throat.
For   
\eqn\twopartclosure{
\cosh (d/l) <   f_c (m, \tilde m)   , \quad {\rm Closed \; Space}
 }
 the space is closed  with an additional image mass, $m^{\prime}$, appearing.  The total mass, $M$,  is negative with  $m^{\prime} = m_{total}= {1\over 4G} - ({-M\over 2G})^{1/2}$.

For $m, \tilde m > m_c/2$, there are only two  kinds of solutions. For  
  $d$ in the range 
\eqn\twopartmergedb{
\cosh (d/l)> f_c (m, \tilde m)      \quad    {\rm Black \; Hole} 
}
the two particles are enclosed by a horizon.
If  $\cosh (d/l)=  f_c (m, \tilde m)      $,  the infinite throat forms at infinity.
For $d$ in the range,  
{\twopartclosure}, 
 the space  is closed with an additional image mass.
Since $f_c (m, \tilde m)     \geq 1$ and  from 
{\twopartclosure}, we recover that   the space is closed in the $l\ra \infty$  flat space limit. 

\subsec{A System Consisting of One Particle and One Black Hole}

 Consider a system consisting of a particle of mass $m$ initially at 
rest  located a geodesic distance $R$ from the horizon 
 of a   black hole of mass $M$ also 
initially at rest. The spatial geometry can again be obtained
from a quotient of the Poincare disk.
  Consider the fundamental region on the Poincare disk associated with the one-black hole solution. This is the region   bound by  two geodesics which do not intersect and are not tangential to one another at infinity.
 Now insert a point particle of mass $m$   a distance $R$ from the horizon of the black hole by excising  a wedge of deficit angle $8\pi G m$ with
vertex at the point particle and   identifying  the two edges. 

The total mass of the system can be found from the effective
one-body generator $\Gamma$ obtained from the   composition of the 
generators  for the black hole
{\blackholemat} and the inverse of the generator  for the translated
 particle
{\translatedparticle}:
\eqn\partbhholonomy{
\Gamma = \Gamma_M ( T_R \G_m T_R^{-1} )^{-1} 
.}
The inverse is taken because the wedge is being {\it removed} from
 the space.
From  the trace   {\totalmass}, we find the total
mass $M_{total}$  of the system 
 \eqn\partbhmass
{\cosh (\pi (8GM_{total})^{1/2} )=     \cosh \b_M  \cos  (4\pi Gm)  +
\sinh (R/l) \;\sinh \b_M \;\sin (4\pi Gm)   }
where $\b_M \equiv \pi (8GM)^{1/2}$.
 As $m\ra 0$, we recover $M_{total} \ra M$, and similarly, as $M\ra 
-1/8G$, we recover $M_{total} =  - (1-4Gm)^2/ (8G)$. 

Depending on the values of the parameters $(m, M ,R)$,  the space 
exterior to the particle and black hole can be open
without a horizon, open with the   particle and black hole  enclosed by a 
horizon, or closed with an additional  image mass.
Consider increasing values of $m$. 

For $0 < \tan (4\pi G m)  < \sinh  \b_M $,    the  different kinds
 of solutions
that can occur as $R$ 
 decreases are as
follows. 
For $R$ in the range
\eqn\mixedopen{
\sinh (R/l) >   {\tan (4\pi G m ) \over \tanh   \b_M   }, {\rm Open, \; No \;Horizon  }
 }
the exterior space is open with total mass  {\partbhmass}, and without any additional horizons.  Again, the total mass increases  with $R$ 
and with 
$m$.  
For $R$ in the range 
\eqn\mixedblackhole{
\sinh (R/l) \leq   {\tan (4\pi G m ) \over \tanh  \b_M }, 
\quad {\rm     Black \; Hole  },
}
the particle and black hole are surrounded by a horizon    with mass  
given by {\partbhmass} as can be verified  directly from
the geometry. 

For $m$ in the range, $ \sinh \b_M \leq   \tan (4\pi G m)  <\infty$, 
{\mixedopen} continues to hold. 
A horizon forms for 
\eqn\mixedblackholeb{ 
{ 1 -      \cosh \b_M  \cos  (4\pi Gm)  \over 
 \sinh \b_M  \;\sin (4\pi Gm) }  
 \leq
\sinh (R/l) \leq   {\tan (4\pi G m ) \over \tanh  \b_M }, 
\quad {\rm     Black \; Hole  }.
}
For $R$ in the range
\eqn\mixedclose{
\sinh (R/l) <   { 1 -      \cosh \b_M  \cos  (4\pi Gm)  \over 
 \sinh \b_M  \;\sin (4\pi Gm) }  \quad {\rm Closed\;Space},
}
the space exterior to the particle and black hole closes with an additional image mass appearing.

Finally, for $m > m_c/2$, $( \tan (4\pi G m ) <0 )$ an open solution without a horizon is not possible. 
In addition,
if $R$ is small enough
\eqn\mixeddegenerate{
\sinh (R/l) <   {\tanh  \b_M \over\tan (4\pi G m )  },     \quad {\rm No\; Solution},
}
instead of the space closing, 
there is no solution.

We now consider solutions describing a point particle   in the   the $M=0$ black hole vacuum {\mzero}.
 These solutions can be obtained from the solutions above by taking a limit in which $M\ra 0$ and $R\ra \infty$ simultaneously.
The distance $R$ to the horizon of the $M>0$ black hole 
from the particle at coordinate $r$ is given by $R = l\cosh^{-1} ( r/r_H) .$  
Now substituting this into the expression {\partbhmass} and taking
$M\ra 0$ yields 
  the total mass of the system 
\eqn\partvacmass{
{\cosh (\pi  (8GM_{total})^{1/2})}=      \cos  (4\pi Gm)  +
\pi r/l \;\sin (4\pi Gm)  
  }
As $m\ra 0$ we recover $M_{total} =0$. We find similar qualitative behavior to the finite $M$ case. As $m$ increases,
a horizon forms at  
\eqn\vaccrit{
\tan (4\pi G m) = \pi r/l
}
For a particle far down the $M=0$ throat $( r <<l)$, it requires an 
arbitrarily small amount of mass    to form a horizon. 
For a particle far away, $r>>l$, a horizon forms
when  $m \approx m_c/2$.

 {\subsec{Two Black Hole Solution}}
Consider   two black holes of masses $M_1$ and $M_2$  initially at rest.
 Let the geodesic 
distance between the horizons  be given by    $d$. 
Depending on the values of the parameters $(M_1, M_2, d)$, 
  either  the  black holes are   enclosed by a third
horizon, or the space is  closed with an additional  image mass.
For $d$ in the range 
\eqn\bhbhbh{
\cosh (d/l) >   g_c^+ (M_1, M_2)  \quad {\rm Black \; Hole},
}
with
\eqn\gdef{
g_c^{\pm} (M_1, M_2)  \equiv {    \cosh \b_{M_{1}} \cosh \b_{M_{2}}  \pm 1 \over 
 \sinh \b_{M_{1}}  \sinh \b_{M_{2} }   }  , \quad
\b_{M_{1,2}} \equiv  \pi (8GM_{1,2})^{1/2} ,  
}
the two black holes are merged, {\it i.e.} surrounded by a  horizon.
The  length of the horizon  $l_0$    is determined from the geometry as follows.
By cutting the space in half along geodesic seams connecting the horizons, one obtains two hexagonal regions with interior right angles. Using hyperbolic geometry, one finds
\eqn\bhbhmass{
\cosh  (l_0/(2l))= \cosh (d/l) \sinh \b_{M_{1}}\sinh \b_{M_{2} }  -  \cosh \b_{M_{1}} \cosh \b_{M_{2 }  }.
} 
For  $\cosh (d/l) =   g_c^+ (M_1, M_2)$, {\bhbhmass} yields $l_0 =0$,
and the infinite $M=0$ throat forms at infinity.   For $d$ in the range 
\eqn\bhbhclosed{
g_c^- (M_1, M_2)  <\cosh (d/l) < g_c^+ (M_1, M_2)  
    \quad {\rm Closed\; Space},
}
the space exterior to the two black holes is closed with an additional  image mass. 
Finally, for 
 $d$ in the range  
 \eqn\bhbhnosolutions{
\cosh (d/l )\leq g_c^- (M_1, M_2)  
   \; \quad {\rm {No \; Solutions}},
}
there are no solutions. 
For the case of equal masses, the right hand side of {\bhbhnosolutions} is 
unity 
and therefore, there are always solutions.
In conclusion, we find 
the striking feature  that when the space exterior to the two black holes
is open, the black holes are necessarily merged {\it i.e.} surrounded
by an apparent horizon.   However, if the space exterior to the two black holes
is    closed,  the  black holes can exist separately
provided there is an additional point mass or a third black hole.
This can be seen from the fact that the solution describing two black holes surrounded by a horizon can alternatively be viewed as a closed universe with three 
separate black holes.
 One can also find solutions in which one or both of the black holes is
the $M=0$ black hole vacuum.
As   in the previous section, these solutions can be obtained by taking a simultaneous
limit in which $M_{1,2}\ra 0$ and $ d\ra \infty$. One finds the same qualitative features as
for non-zero mass.  
    
\newsec{Conclusion}

The next step is to evolve this initial data.  An exact time-dependent solution describing the 
merging of two bodies to form a black hole should be possible. It would also be 
interesting to see whether critical behavior of the sort found by Choptuik
{\ref\choptuik{ M. Choptuik, {\it Phys. Rev. Lett. } {\bf 70} (1993) 9.}}
exists in this case.
Since   previous studies of critical behavior in gravitational collapse
were restricted to   cases
with a high degree of symmetry, this would
be an example of critical behaviour under more general circumstances.     
  In addition, the collison of two bodies  likely produces a naked singularity for
certain initial conditions.
This  is reminiscient of {\ref\kt{D. Kastor and J.
 Traschen, {\it Phys. Rev. D} {\bf 47} (1993) 5370.}}{ \ref\bhkt{ D. Brill, G. 
Horowitz, D. Kastor, and J. Traschen, {\it Phys. Rev. D} {\bf 49} (1994) 840.}}
 where it was shown that extremal charged black holes
in four dimensions  in a theory with cosmological constant can 
collide and form naked singularities. Also, the extremal $M=0$ black hole 
vacuum
  is similar in some ways to the extremal charged $3+1$
 dimensional black hole.

 \vskip 20pt

\centerline{\bf Acknowledgements}

I  would like to thank Steve Carlip, Richard Epp, Joel Hass, and  Yoav Peleg  
 for helpful discussions and and  for useful
comments on the paper.
This work was supported by  NSF grant  PHY-93-57203.

Note Added:
Upon completion of  this paper,  I received a draft of a paper from  Dieter Brill  
(gr-qc/9511022)  which included a discussion of initial data for multi-black hole solutions.
\baselineskip=30pt

\listrefs
\end